\documentclass[prl,twocolumn,floatfix,amsmath,amssymb]{revtex4}
\pdfoutput=1
\usepackage{graphicx}
\usepackage{dcolumn}
\usepackage{bm}
\usepackage{hyperref}
\begin{document}
\title{High temperature superconductivity: from complexity to simplicity}
\author{Sudip Chakravarty}
\affiliation{Department of Physics and Astronomy, University of
California Los Angeles, Los Angeles, CA 90095, USA}

\date{\today}

\begin{abstract}
I discuss the recent quantum oscillation experiments in the underdoped high temperature superconductors.
\end{abstract}
\maketitle

\begin{center}
\begin{quote}HAMLET: Do you see yonder cloud that's almost in shape of a camel ?\\
POLONIUS: By th' mass and 'tis, like a camel indeed.\\
HAMLET: Methinks it is like a weasel.\\
POLONIUS: It is backed like a weasel.\\
HAMLET: Or like a whale.\\
POLONIUS: Very like a whale.\\
---{\em William Shakespeare}
\end{quote}
\end{center}

More than 20 years ago, Bednorz and
M\"uller discovered superconductivity
in copper oxides at remarkably
high temperatures~\cite{Bednorz:1986}. Since then, physicists
have struggled to understand the mechanisms
at work. Recently, a set of experiments on
cuprates in high magnetic fields ~\cite{Doiron-Leyraud:2007,Bangura:2007,LeBoeuf:2007,Yelland:2007,Jaudet:2007} has
completely changed the landscape of research
in high-temperature superconductors (HTSs).
In particular, the data suggest that the current
carriers are both electrons and holes, when in
fact the materials are Òhole dopedÓ---i.e., the
current carriers should be positively charged.
Moreover, the data cannot be reconciled with
an important theorem about how electrons are
organized in materials ~\cite{Luttinger:1960} unless one assumes
that the signals arise from a combination of
both holes and electrons. Until now, physicists
have not been able to decide whether the
cuprates, in Shakespeare's terms, are camels
or whales; in fact, these experiments foreshadow
a remarkable degree of simplicity in
these complex materials.

The cuprates start out as insulators and
become superconductors when doped with
additional charge carriers. These so-called
Mott insulators insulate by virtue of strong
repulsive Coulomb interaction and need not
break any symmetries in the lowest energy
state, the ground state. A symmetry of a system
is a transformation, such as a translation or a
rotation, that keeps it unchanged. Such a 
symmetry
is said to be broken, or spontaneously
broken, if the system does not obey the symmetry
of the underlying fundamental physical
nature of the material; for example, a ferromagnet
breaks the spin-rotational symmetry with
its magnetization pointing in a definite direction.
The notion of symmetry and broken symmetry
finds many deep applications in physics.

Soon after the discovery of the cuprate
superconductors, Anderson proposed ~\cite{Anderson:1987} that
their parent compounds begin as a featureless
spin liquid that does not break any symmetries,
called the resonating valence bond
(RVB) state: ``The preexisting magnetic
singlet pairs of the insulating state become
charged superconducting pairs when the insu
lator is doped sufficiently strongly''~\cite{Anderson:1987}.
Unfortunately, experiments show that the
insulating phase is a simple antiferromagnet
in which the spins are arranged in antiparallel
manner, that is, with a broken symmetry. The
materials remain antiferromagnets for a range
of doping, and then, after a sequence of not
well understood states as a function of doping,
they become superconductors.

How this plays out experimentally can be
understood by looking at the Fermi surface, a
fundamental concept in condensed matter
physics. The Fermi surface differentiates the
occupied electronic states from the unoccupied
states (in coordinates of momentum
rather than ÒrealÓ space). Electrons fill the
Fermi surface (FS) up to some highest occupied
energy called the Fermi energy (see the figure).
The excitations from the FS
(e.g., when a current flows) are called Landau
quasiparticles (quasi, because they are combinations
or superpositions of real particles).
The  robustness of FS is due to its topological invariance, one of the most basic invariances in mathematical physics, which signifies stability with respect  to ``small deformations''~\cite{Volovik:2003}. Even when the quasiparticles are absent due to electron-electron interactions, as in one-dimensional electronic systems, the FS is still defined by the same topological invariant. A reconstruction of this surface, such as a break up of a single surface into hole-like and electron-like pockets, requires a global deformation, most likely  a broken symmetry. 
The new experimental work (2Ð6) yielded
measurements of the oscillations that arise
from energy levels created by imposing a
magnetic field on the material (the Landau
levels). As the magnetic field is increased, the
highest fully occupied levels sweep past the
Fermi energy, and the system periodically
returns to itself, hence the oscillation in physical
properties. The oscillations of the Hall
resistance ~\cite{Doiron-Leyraud:2007,LeBoeuf:2007}, capable of detecting the sign
of the charge carriers, seem to show the presence
of electron and hole pockets in the Fermi
surface, suggesting that it undergoes some
kind of reconstruction. 

One might complain that these high field
measurements are still considerably below
the upper critical field where superconductivity
disappears (about 100 T or more) and
are affected by the complex motion of vortices
generated by the magnetic field. This
may be true, but quantum oscillations in
many superconductors are observed at fields
as small as half the critical field, with the
oscillation frequencies unchanged from the
nonsuperconducting state (with an increased
damping, however). It is also known that the
quasiparticles of HTSs do not form Landau
levels (9). Thus, it is very likely that the quantum
oscillation experiments are accessing the
normal state beyond the realm of superconductivity.
But what kind of state? As the oscillations
definitively point to both electron and
hole pockets, it cannot be a conventional
Fermi surface, rather one that has undergone
a reconstruction due to a broken symmetry
at variance with the RVB picture~\cite{Chakravarty:2007}.

We may be finally beginning to understand
these superconductors after two decades. The
fly in the ointment is the lack of observation of
electron and hole pockets in other measurements
in hole-doped superconductors (in angle resolved
photoemission spectroscopy, for instance)
that are also capable of measuring
Fermi surfaces [see, however, the work on electron-doped materials~\cite{Armitage:2001}]. Missing so far in
experiments are also the higher frequency
oscillations that must arise from the hole pockets,
not just the electron pockets ~\cite{LeBoeuf:2007}. With further
experimental work, we should be able to tell
just what kind of animal we are dealing with.
 [{\em Note added\;}: A higher frequency is now seen and is believed to arise from an incommensurate order---S. Sebastian, N. Harrison, and G. Lonzarich, private communication.]

This work was supported by NSF under grant:  DMR-0705092. Special thanks are due to Hae-Young Kee for many discussions and collaborations. I would also like to thank E. Abrahams, N. P. Armitage, R. B. Laughlin, Z. Tesanovic, and J. Zaanen for important comments.

\begin{figure}[htb]
\caption{{\bf Fermi surface reconstruction:} In a crystalline lattice of periodicity $a$, the available quantum states are contained within the Brillouin zone (BZ). In ({\bf a}) we have shown a two-dimensional example, $-\frac{\pi}{a}\le k_{x} <\frac{\pi}{a}$, $-\frac{\pi}{a}\le k_{y} <\frac{\pi}{a}$, which is pertinent to high temperature superconductors that consist of weakly coupled CuO-planes. For $N$-atoms, with unit cells containing 1 atom in a square $a\times a$,  there are exactly $2N$ available states, as each state can be filled by electrons with two distinct spin orientations. The diamond in ({\bf a}) is called the reduced Brillouin zone (RBZ) and contains exactly half the number of available states.  In ({\bf a}) the unoccupied states are colored red; these, holes, in a otherwise filled BZ, respond as charge carriers with positive sign in response to electric fields. The constant energy contours are shown as the set of black curves. The filled diamond corresponds to one electron per unit cell. The complementary sets within the BZ,  when reassembled together, will form an identical diamond, which we may say is filled with holes.  In ({\bf a})  the red area corresponds to $(1+x)$ holes per unit cell of the crystal lattice. The excess, $x$, is called the doped holes.
High temperature superconductors are schizophrenic. In some regimes they behave as though they consist of $(1+x)$ charge carriers, holes, while in the  underdoped regime their properties are determined instead by $x$ doped holes. This is an important mystery. A class of theories posit that FS reconstructs in the uderdoped regime. Consider shifting the FS in ({\bf a}) by vectors $(\pm \frac{\pi}{a},\pm \frac{\pi}{a})$, which will give rise to the Figure ({\bf b}), ignoring the shading for clarity. Interesting quantum mechanical  processes, about which there can be much debate, can result in reconnections shown in ({\bf c}), as in a kaleidoscope. However, if we continued to consider the full BZ, we would double the number of states. All distinct states are contained  in the RBZ, but there are now two distinct sets of energy levels, the upper band and the lower band. However, we continue to use the full  BZ in ({\bf c}), as a better aid for visualization. Because the RBZ is the fundamental unit in the wave vector space, the new unit cell of the crystal lattice is doubled, given by a square $\sqrt{2}a\times\sqrt{2}a$, and the full translational symmetry of the original lattice is broken. The FS now consists of {\em disconnected sheets} of blue and red areas. The remarkable fact is that the charge carriers in the blue region behave like electrons of fraction $n_{e}$ and in the red region like holes  of fraction $n_{h}$. The doped holes is easily seen  to be $x=2n_{h}-n_{e}$, as there are two hole pockets and one electron pocket in the RBZ. The broken symmetry invoked here is called commensurate, as the translational invariance of the crystal of integer multiples of the next nearest neighbor  lattice vectors  of the original lattice is still preserved. The broken symmetry can also be incommensurate with the original crystal lattice and can give rise to more complex FS reconstructions. }
\end{figure}

\clearpage

\begin{figure}[htb]
\begin{center}
\includegraphics[scale=0.8]{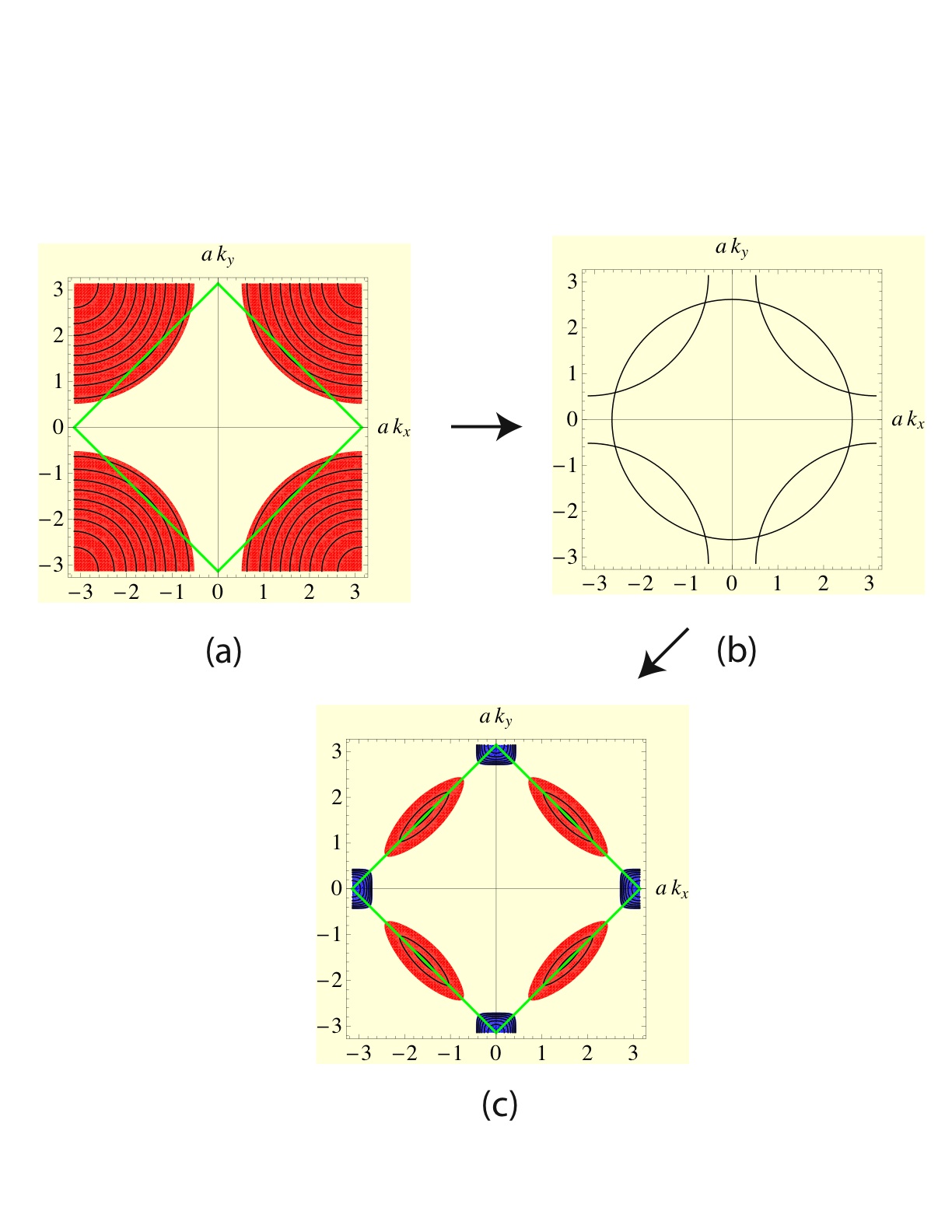}
\end{center}
\end{figure}


\begin{thebibliography}{10}
\expandafter\ifx\csname natexlab\endcsname\relax\def\natexlab#1{#1}\fi
\expandafter\ifx\csname bibnamefont\endcsname\relax
  \def\bibnamefont#1{#1}\fi
\expandafter\ifx\csname bibfnamefont\endcsname\relax
  \def\bibfnamefont#1{#1}\fi
\expandafter\ifx\csname citenamefont\endcsname\relax
  \def\citenamefont#1{#1}\fi
\expandafter\ifx\csname url\endcsname\relax
  \def\url#1{\texttt{#1}}\fi
\expandafter\ifx\csname urlprefix\endcsname\relax\def\urlprefix{URL }\fi
\providecommand{\bibinfo}[2]{#2}
\providecommand{\eprint}[2][]{\url{#2}}

\bibitem{Bednorz:1986}J. G. Bednorz and K. A. M\"uller, Z. Physik B {\bf 64}, 189 (1986).

\bibitem[{\citenamefont{Doiron-Leyraud
  et~al.}(2007)\citenamefont{Doiron-Leyraud, Proust, LeBoeuf, Levallois,
  Bonnemaison, Liang, Bonn, Hardy, and Taillefer}}]{Doiron-Leyraud:2007}
\bibinfo{author}{\bibfnamefont{N.}~\bibnamefont{Doiron-Leyraud}},
  \bibinfo{author}{\bibfnamefont{C.}~\bibnamefont{Proust}},
  \bibinfo{author}{\bibfnamefont{D.}~\bibnamefont{LeBoeuf}},
  \bibinfo{author}{\bibfnamefont{J.}~\bibnamefont{Levallois}},
  \bibinfo{author}{\bibfnamefont{J.~B.} \bibnamefont{Bonnemaison}},
  \bibinfo{author}{\bibfnamefont{R.~X.} \bibnamefont{Liang}},
  \bibinfo{author}{\bibfnamefont{D.~A.} \bibnamefont{Bonn}},
  \bibinfo{author}{\bibfnamefont{W.~N.} \bibnamefont{Hardy}}, \bibnamefont{and}
  \bibinfo{author}{\bibfnamefont{L.}~\bibnamefont{Taillefer}},
  \bibinfo{journal}{Nature} \textbf{\bibinfo{volume}{447}},
  \bibinfo{pages}{565} (\bibinfo{year}{2007}).

\bibitem[{\citenamefont{Bangura et~al.}(2007)\citenamefont{Bangura, Fletcher,
  Carrington, Levallois, Nardone, Vignolle, Heard, Doiron-Leyraud, LeBoeuf,
  Taillefer et~al.}}]{Bangura:2007}
\bibinfo{author}{\bibfnamefont{A.~F.} \bibnamefont{Bangura}},
  \bibinfo{author}{\bibfnamefont{J.~D.} \bibnamefont{Fletcher}},
  \bibinfo{author}{\bibfnamefont{A.}~\bibnamefont{Carrington}},
  \bibinfo{author}{\bibfnamefont{J.}~\bibnamefont{Levallois}},
  \bibinfo{author}{\bibfnamefont{M.}~\bibnamefont{Nardone}},
  \bibinfo{author}{\bibfnamefont{B.}~\bibnamefont{Vignolle}},
  \bibinfo{author}{\bibfnamefont{P.~J.} \bibnamefont{Heard}},
  \bibinfo{author}{\bibfnamefont{N.}~\bibnamefont{Doiron-Leyraud}},
  \bibinfo{author}{\bibfnamefont{D.}~\bibnamefont{LeBoeuf}},
  \bibinfo{author}{\bibfnamefont{L.}~\bibnamefont{Taillefer}},
  \bibnamefont{et~al.}, \bibinfo{journal}{arXiv:0707.4461}
  (\bibinfo{year}{2007}).

\bibitem[{\citenamefont{LeBoeuf~{\em et al.}}(2007)}]{LeBoeuf:2007}
\bibinfo{author}{\bibfnamefont{D.}~\bibnamefont{LeBoeuf~{\em et al.}}},
  \bibinfo{journal}{Nature} \textbf{\bibinfo{volume}{450}},
  \bibinfo{pages}{533} (\bibinfo{year}{2007}).

\bibitem[{\citenamefont{Yelland~{\em et al.}}(2007)}]{Yelland:2007}
\bibinfo{author}{\bibfnamefont{D.~A.} \bibnamefont{Yelland~{\em et al.}}},
  \bibinfo{journal}{arXiv:0707.0057}  (\bibinfo{year}{2007}).

\bibitem[{\citenamefont{Jaudet et~al.}(2007)\citenamefont{Jaudet, Vignolles,
  Audouard, Levallois, LeBoeuf, Doiron-Leyraud, Vignolle, Nardone, Zitouni,
  Liang et~al.}}]{Jaudet:2007}
\bibinfo{author}{\bibfnamefont{C.}~\bibnamefont{Jaudet}},
  \bibinfo{author}{\bibfnamefont{D.}~\bibnamefont{Vignolles}},
  \bibinfo{author}{\bibfnamefont{A.}~\bibnamefont{Audouard}},
  \bibinfo{author}{\bibfnamefont{J.}~\bibnamefont{Levallois}},
  \bibinfo{author}{\bibfnamefont{D.}~\bibnamefont{LeBoeuf}},
  \bibinfo{author}{\bibfnamefont{N.}~\bibnamefont{Doiron-Leyraud}},
  \bibinfo{author}{\bibfnamefont{B.}~\bibnamefont{Vignolle}},
  \bibinfo{author}{\bibfnamefont{M.}~\bibnamefont{Nardone}},
  \bibinfo{author}{\bibfnamefont{A.}~\bibnamefont{Zitouni}},
  \bibinfo{author}{\bibfnamefont{R.}~\bibnamefont{Liang}},
  \bibnamefont{et~al.}, \bibinfo{journal}{arXiv.org:0711.3559}
  (\bibinfo{year}{2007}).

\bibitem[{\citenamefont{Luttinger}(1960)}]{Luttinger:1960}
\bibinfo{author}{\bibfnamefont{J.~M.} \bibnamefont{Luttinger}},
  \bibinfo{journal}{Phys. Rev.} \textbf{\bibinfo{volume}{119}},
  \bibinfo{pages}{1153} (\bibinfo{year}{1960}).

\bibitem[{\citenamefont{Anderson}(1987)}]{Anderson:1987}
\bibinfo{author}{\bibfnamefont{P.~W.} \bibnamefont{Anderson}},
  \bibinfo{journal}{Science} \textbf{\bibinfo{volume}{235}},
  \bibinfo{pages}{1196} (\bibinfo{year}{1987}).
  
  \bibitem{Volovik:2003}G. E. Volovik, {\em The universe in a Helium droplet} (Cambridge University Press, Cambridge, 2003)

\bibitem[{\citenamefont{Franz and Tesanovic}(2000)}]{Franz:2000}
\bibinfo{author}{\bibfnamefont{M.}~\bibnamefont{Franz}} \bibnamefont{and}
  \bibinfo{author}{\bibfnamefont{Z.}~\bibnamefont{Tesanovic}},
  \bibinfo{journal}{Phys. Rev. Lett.} \textbf{\bibinfo{volume}{84}},
  \bibinfo{pages}{554} (\bibinfo{year}{2000}).

\bibitem[{\citenamefont{Chakravarty and Kee}(2007)}]{Chakravarty:2007}
\bibinfo{author}{\bibfnamefont{S.}~\bibnamefont{Chakravarty}} \bibnamefont{and}
  \bibinfo{author}{\bibfnamefont{H.-Y.} \bibnamefont{Kee}},
  \bibinfo{journal}{arXiv:0710.0608}  (\bibinfo{year}{2007}).

\bibitem[{\citenamefont{Armitage et~al.}(2001)\citenamefont{Armitage, Lu, Kim,
  Damascelli, Shen, Ronning, Feng, Bogdanov, Shen, Onose
  et~al.}}]{Armitage:2001}
\bibinfo{author}{\bibfnamefont{N.~P.} \bibnamefont{Armitage}},
  \bibinfo{author}{\bibfnamefont{D.~H.} \bibnamefont{Lu}},
  \bibinfo{author}{\bibfnamefont{C.}~\bibnamefont{Kim}},
  \bibinfo{author}{\bibfnamefont{A.}~\bibnamefont{Damascelli}},
  \bibinfo{author}{\bibfnamefont{K.~M.} \bibnamefont{Shen}},
  \bibinfo{author}{\bibfnamefont{F.}~\bibnamefont{Ronning}},
  \bibinfo{author}{\bibfnamefont{D.~L.} \bibnamefont{Feng}},
  \bibinfo{author}{\bibfnamefont{P.}~\bibnamefont{Bogdanov}},
  \bibinfo{author}{\bibfnamefont{Z.-X.} \bibnamefont{Shen}},
  \bibinfo{author}{\bibfnamefont{Y.}~\bibnamefont{Onose}},
  \bibnamefont{et~al.}, \bibinfo{journal}{Phys. Rev. Lett.}
  \textbf{\bibinfo{volume}{87}}, \bibinfo{pages}{147003}
  (\bibinfo{year}{2001}).

\end{thebibliography}
\end{document}